%% file: main.tex
\documentclass[sigconf]{acmart} 

\usepackage{multirow} 
\usepackage{subcaption}
\usepackage{enumitem}
\usepackage{booktabs, tabularx}
\usepackage{xcolor}
\usepackage{makecell} 
\usepackage{xurl}
\usepackage{arydshln}
\usepackage{tcolorbox}
\usepackage[symbol]{footmisc}
\usepackage{nicematrix}
\usepackage{booktabs}
\usepackage{tikz}

\newcommand{\updated}[1]{\textcolor{black}{#1}}

\newcommand{\ready}[1]{\textcolor{black}{#1}}

\AtBeginDocument{
  }

\copyrightyear{2026}
\acmYear{2026}
\setcopyright{cc}
\setcctype{by}
\acmConference[SIGIR '26]{Proceedings of the 49th International ACM SIGIR Conference on Research and Development in Information Retrieval}{July 20--24, 2026}{Melbourne, VIC, Australia}
\acmBooktitle{Proceedings of the 49th International ACM SIGIR Conference on Research and Development in Information Retrieval (SIGIR '26), July 20--24, 2026, Melbourne, VIC, Australia}
\acmDOI{10.1145/3805712.3808558}
\acmISBN{979-8-4007-2599-9/2026/07}

\begin{document}


\input{title}
\input{chapter1.tex}
\input{chapter2.tex}
\input{chapter3.tex}

\input{chapter4.tex} 
\input{chapter5.tex}
\input{references}

\end{document}

%% file: title.tex
\title{Where Relevance Emerges: A Layer-Wise Study \\ of Internal Attention for Zero-Shot Re-Ranking}

\author{Haodong Chen}
\affiliation{
  \institution{The University of Queensland}
  \city{Brisbane}
  \country{Australia}
}
\email{haodong.chen1@student.uq.edu.au}

\author{Shengyao Zhuang}
\affiliation{
  \institution{The University of Queensland}
  \city{Brisbane}
  \country{Australia}
}
\email{s.zhuang@uq.edu.au}

\author{Zheng Yao}
\affiliation{
  \institution{The University of Queensland}
  \city{Brisbane}
  \country{Australia}
}
\email{zheng.yao1@student.uq.edu.au}

\author{Guido Zuccon}
\affiliation{%
  \institution{The University of Queensland}
  \city{Brisbane}
  \country{Australia}
}
\email{g.zuccon@uq.edu.au}

\author{Teerapong Leelanupab}
\affiliation{%
	\institution{The University of Queensland}
	\city{Brisbane}
	\country{Australia}
}
\email{t.leelanupab@uq.edu.au}


\begin{abstract}
Zero-shot document re-ranking with Large Language Models (LLMs) has evolved from \textit{Pointwise} methods to \textit{Listwise} and \textit{Setwise} approaches that optimize computational efficiency. Despite their success, these methods predominantly rely on generative scoring or output logits, which face bottlenecks in inference latency and result consistency. \textit{In-Context Re-ranking} (ICR) has recently been proposed as an $O(1)$ alternative method. ICR extracts internal attention signals directly, avoiding the overhead of text generation. However, existing ICR methods simply aggregate signals across all layers; layer-wise contributions and their consistency across architectures have been left unexplored. Furthermore, no unified study has compared internal attention with traditional generative and likelihood-based mechanisms across diverse ranking frameworks under consistent conditions.

In this paper, we conduct an orthogonal evaluation of \textit{generation}, \textit{likelihood}, and \textit{internal attention} mechanisms  across multiple ranking frameworks. We further identify a universal ``bell-curve'' distribution of relevance signals across transformer layers, which motivates the proposed \textit{Selective-ICR} strategy that reduces inference latency by 30\%--50\% without compromising effectiveness. \updated{Finally, evaluation on the reasoning-intensive BRIGHT benchmark shows that precisely capturing high-quality in-context attention signals fundamentally reduces the need for model scaling and reinforcement learning: a zero-shot 8B model matches the performance of 14B reinforcement-learned re-rankers, while even a 0.6B model outperforms state-of-the-art generation-based approaches.} These findings redefine the efficiency-effectiveness frontier for LLM-based re-ranking and highlight the latent potential of internal signals for complex reasoning ranking tasks. Our code and results are publicly available at \textcolor{blue}{\url{https://github.com/ielab/Selective-ICR}}.

\end{abstract}


\maketitle

%% file: chapter1.tex
\section{Introduction and Related Work}
\label{intro}
Zero-shot document re-rankers based on Large Language Models (LLMs)~\cite{ma:ZeroShotListwiseDocument:2023,pradeep:RankVicunaZeroShotListwise:2023,pradeep:RankZephyrEffectiveRobust:2023,zhuang:SetwiseApproachEffective:2024,qin:LargeLanguageModels:2024,zhuang:RankR1EnhancingReasoning:2025} have emerged as a high-performance alternative to traditional neural ranking methods~\cite{lin:PretrainedTransformersText:2021}, eliminating the need for task-specific fine-tuning. These LLM-based zero-shot rankers typically employ diverse prompting strategies, which can be broadly categorized into four main approaches. \textit{Pointwise} methods instruct the LLM to evaluate each candidate document independently, usually by generating a relevance score or a binary label for a single query-document pair~\cite{liang:HolisticEvaluationLanguage:2023,sachan:ImprovingPassageRetrieval:2023,zhuang:OpensourceLargeLanguage:2023}. Moving beyond independent assessment, \textit{Pairwise} approaches instead model the relative interactions between document pairs~\cite{qin:LargeLanguageModels:2024}, and \textit{Listwise} strategies process an entire set of candidates simultaneously to output a direct permutation of identifiers~\cite{ma:ZeroShotListwiseDocument:2023,sun:ChatGPTGoodSearch:2023,pradeep:RankVicunaZeroShotListwise:2023}. While both seek to capture inter-document dependencies, they are often limited by either the high computational cost of exhaustive comparisons or the context constraints of a single prompt. To bridge this gap, the \textit{Setwise} approach instructs the LLM to select only the most relevant document from a given candidate set at each step~\cite{zhuang:SetwiseApproachEffective:2024}. This approach allows sorting algorithms such as \textit{heapsort} and \textit{bubblesort} to iteratively infer preferences among multiple documents, thereby significantly reducing the total number of required comparisons and leading to substantial computational savings.

Despite their success, existing re-ranking methods predominantly rely on \textit{generative} capabilities, either by prompting for formatted labels or utilizing output logits to bypass autoregressive decoding~\cite{reddy:FIRSTFasterImproved:2024,zhuang:DeepQueryLikelihood:2021,zhuang:TILDETermIndependent:2021}. However, this reliance remains constrained by the high cost of multiple inference forward passes and the risk of output instability.

To address these limitations, \textit{In-Context Re-ranking} (ICR) \cite{chen:AttentionLargeLanguage:2025} was proposed as an alternative approach that avoids the overhead of text generation by extracting relevance signals directly from the model's internal attention patterns. ICR assumes that more relevant documents should receive greater attention weights when an LLM processes the query tokens. Accordingly, it calculates ranking scores by aggregating the attention weights received by document tokens from all query tokens across all layers and attention heads. Since the attention weights of each query token form a normalized distribution summing to one, this method prevents longer documents from receiving inflated scores simply due to their higher token count. To eliminate intrinsic model biases\textemdash such as preferences for specific token positions or syntactic structures\textemdash ICR employs a dual-pass calibration strategy: it subtracts attention scores obtained using a content-free query (e.g., ``N/A'') from those of the actual search query, yielding semantic relevance scores. By leveraging these semantics-driven attention signals, ICR only requires two forward passes to rank an entire set of $N$ documents, resulting in a constant number of inference passes ($O(1)$) with respect to $N$. This makes it substantially more efficient than generative re-ranking methods, which typically require at least $O(N)$ forward passes.

Despite its efficiency, the internal behavior of ICR remains largely unexplored. The original study by Chen et al. \cite{chen:AttentionLargeLanguage:2025} relied on aggregate signals across all transformer layers and was limited to few models and scales, specifically 7B and 8B parameter families. This leaves a gap in understanding how these signals are distributed across layers. It remains unknown whether these internal signals follow a consistent pattern and, if so, whether such a pattern holds across various model families, parameter scales, and diverse datasets. By addressing this question, we can develop principled layer-selection strategies that improve both effectiveness and efficiency\textemdash aggregating only signal-rich layers and enabling earlier termination of forward passes. Furthermore, there is currently no unified study that provides a side-by-side comparison of internal attention signals against traditional generative and likelihood-based mechanisms under consistent experimental conditions. Consequently, this paper proposes the following research questions:

\vspace{2pt}
\begin{itemize}[leftmargin=24pt, itemsep=2pt]
    \item[\textbf{RQ1:}] How is the relevance signal distributed across the layers of an LLM, and do all layers contribute equally to the effectiveness of \textit{In-Context Re-ranking} (ICR)?
    \item[\textbf{RQ2:}] How does attention-based scoring compare with traditional generative and likelihood-based methods when integrated into \textit{Listwise} and \textit{Setwise} ranking frameworks?
    \item[\textbf{RQ3:}] Does the attention-based ranking approach remain effective when processing \textit{reasoning-intensive} tasks, and is the layer-wise signal distribution universal across varying task complexities and model scales?
\end{itemize}
\vspace{2pt}

To investigate these questions, we perform an extensive evaluation involving over 167,000 query runs across different models and major benchmarks, which lead to the following findings and contributions:

\begin{itemize}[leftmargin=*, itemsep=2pt] 
    \item We demonstrate that relevance signals are non-uniformly distributed across transformer layers. Based on this observation, we propose \textit{Selective-ICR}, a strategy that isolates high-signal layers to optimize ranking effectiveness and computational efficiency. 
    \item We conduct a systematic cross-evaluation of three scoring mechanisms—\textit{generative}, \textit{likelihood}, and \textit{internal attention}—against two ranking strategies, \textit{listwise} and \textit{setwise}, providing the first comprehensive analysis benchmarking these disparate paradigms on a unified process under strictly identical conditions.
    \item We show that isolating a specific \textit{Oracle Layer} enables a zero-shot Llama~3.1~8B model to achieve performance parity with a 14B RL-trained model, while even our smallest 0.6B model outperforms the GPT-4-based Listwise reranker on BRIGHT benchmark~\cite{su:BRIGHTRealisticChallenging:2025}. This indicates that internal attention can effectively resolve complex, reasoning-intensive tasks, showing immense potential.
\end{itemize}

%% file: chapter2.tex
\section{Probing Layer-wise Relevance Signals}
\label{sec:rq1}

Existing In-Context Re-ranking (ICR) work typically aggregates attention weights across all transformer layers under the assumption of uniform attention contribution~\cite{chen:AttentionLargeLanguage:2025}. In this section, we explore \textbf{RQ1} by challenging this assumption through an investigation of the relevance distribution across different transformer layers. Our goal is to determine whether certain layer intervals provide higher-quality signals and whether this pattern remains consistent across varying model scales and diverse tasks.

\begin{figure}[t]
\centering
\begin{subfigure}[b]{\linewidth}
    \centering
    \includegraphics[width=\linewidth]{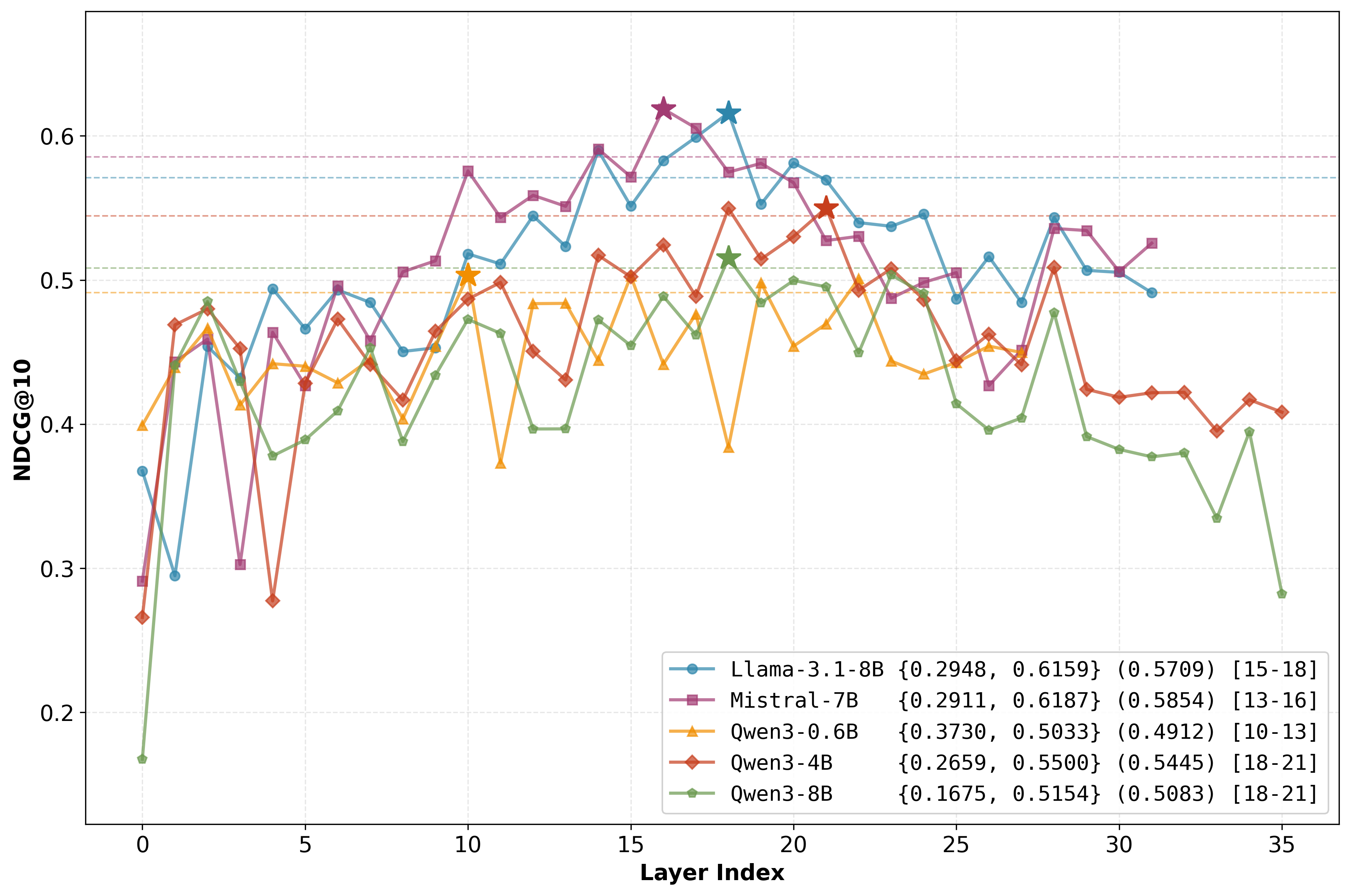}
    \caption{TREC-DL-2019 - Layer-wise Performance Analysis}
    \label{fig:dl19}
\end{subfigure}
\hfill
\vspace{4pt}
\begin{subfigure}[b]{\linewidth}
    \centering
    \includegraphics[width=\linewidth]{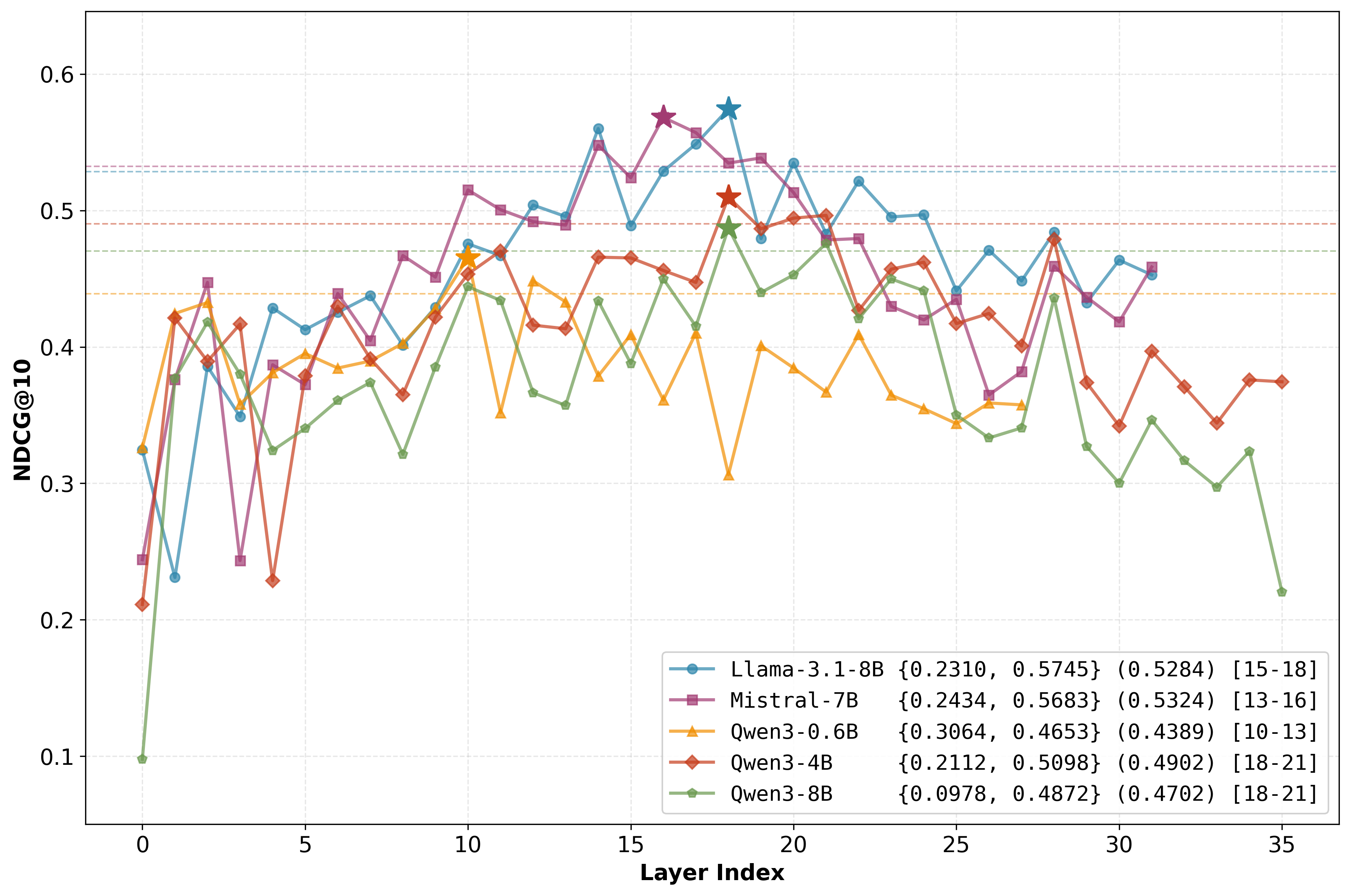}
    \caption{TREC-DL-2020 - Layer-wise Performance Analysis}
    \label{fig:dl20}
\end{subfigure}
\caption{\updated{Layer-wise performance analysis (\textit{nDCG@10}) on TREC-DL 2019 and 2020. Stars ($\star$) indicate the peak performance achieved by a single layer. Curly brackets \{min, max\} denote the performance range across all layers. Dashed horizontal lines and values in parentheses $(\cdot)$ represent the \textit{nDCG@10} obtained from the full-layer aggregation strategy~\cite{chen:AttentionLargeLanguage:2025}. Square brackets $[l_1\!-\!l_2]$ indicate selected layers for the aggregation interval.}}
\label{fig:pilot_study}
\end{figure}

\subsection{Experimental Design}
To analyze the heterogeneous contributions of individual layers, we adopt a two-stage experimental process:
(1) a pilot study conducted on the TREC-DL 2019 and 2020 benchmarks \cite{craswell:OverviewTREC2019:2020, craswell:OverviewTREC2020:2021} to probe the relevance distribution and identify optimal layer intervals; and 
(2) a full-scale evaluation across nine datasets from the BEIR benchmark~\cite{thakur:BEIRHeterogenousBenchmark:2021} to validate the generalizability and efficiency of our proposed \textit{Selective-ICR} strategy.

For this, we evaluate the Mistral-7B~\cite{jiang:Mistral7B:2023} and Llama-3.1-8B~\cite{grattafiori:Llama3Herd:2024} models, following the protocols established by Chen et al.~\cite{chen:AttentionLargeLanguage:2025}. To further investigate the universality of the layer-wise relevance distribution, we extend our evaluation to the Qwen3 series (0.6B, 4B, and 8B)~\cite{yang:Qwen3TechnicalReport:2025}. This selection allows us to examine whether signal patterns scale with model size.

For each query, we re-rank the top 100 candidate documents retrieved by BM25~\cite{robertson:TheProbabilisticRelevance:2009}. To maintain a global attention context and avoid the performance degradation associated with sliding windows \cite{chen:AttentionLargeLanguage:2025}, we process all 100 candidates in a single pass. Each document is truncated to its leading 300 words to balance context coverage with memory constraints. We adopt $nDCG@10$ as our primary metric to ensure comparability with existing zero-shot re-ranking benchmarks.

To investigate the layer-wise contribution to ranking effectiveness, we decompose the original all-layer ICR aggregation into a granular layer-by-layer evaluation. Specifically, we isolate each individual transformer layer and compute independent relevance scores based solely on its internal attention distribution. \updated{Crucially, the performance metrics are calculated per-layer; for instance, the $nDCG@10$ for layer $l$ is derived from its own signals, rather than cumulatively aggregating signals from layer $0$ to layer $|L|$, where $L$ denotes the set of transformer layers.} To ensure a fair comparison, each layer-wise evaluation strictly maintains the same \textit{null-input calibration} and \textit{reversed sequence ordering} to position higher-ranked candidates closer to the query as used in the full-scale implementation. This methodology allows us to map the ``relevance evolution'' as information propagates through the model's depth while isolating the specific impact of each layer's representational capacity.

All tasks were executed on one NVIDIA A100 (80GB) GPU, utilizing \textit{Flash Attention 2} to accelerate the extraction of attention weights and ensure consistent, reproducible latency measurements.

\subsection{Layer-wise Pilot study on TREC-DLs}
We first conduct a layer-wise pilot study on the TREC-DL 2019 and 2020 benchmarks to identify the optimal layer intervals. As shown in Figure~\ref{fig:pilot_study}, the ranking signal exhibits a ``bell-curve'' trend across all tested models. Specifically, the middle layers near the \textit{center} capture deep semantic relevance more effectively than the initial layers, which focus on surface-level features, or the final layers, which are often specialized for next-token prediction~\cite{lad2025remarkable,song2025demystifying,zhang2024investigating}. \updated{Crucially, we observe that the peak performance achieved by a single optimal layer\textemdash indicated by $\star$ and corresponding to the upper bound of the layer-wise range $\{\min,\max\}$\textemdash consistently surpasses the results obtained from aggregating all layers, shown in parentheses $(\cdot)$.}

This observation reveals a significant limitation in the original ICR approach: assuming uniform positive contributions across the entire model depth leads to a ``signal dilution effect'', where strong relevance signals from the middle layers are obscured by noise from suboptimal layers, including surface-level, syntactic, and task-specific output signals.

\subsection{Selective-ICR}
\begin{figure*}[t]
\centering
\includegraphics[width=1.00\textwidth]{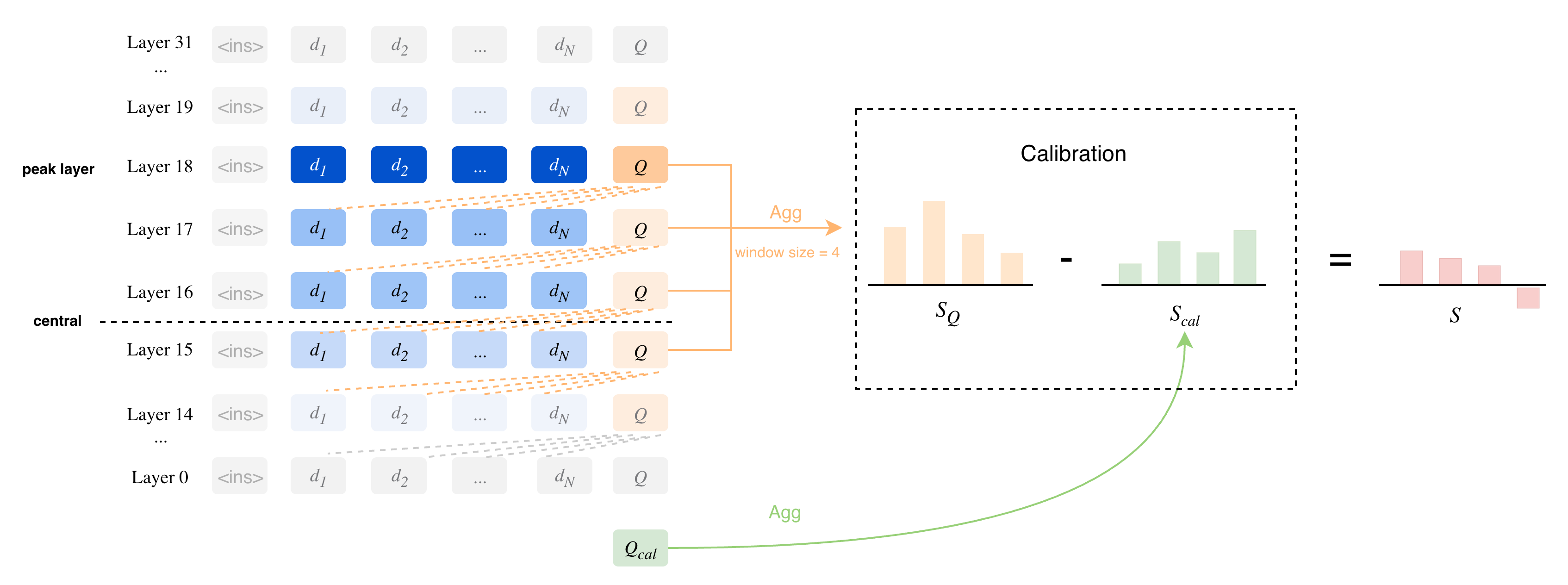}
\caption{Illustration of Selective-ICR with Center-Biased Interval Selection, exemplified by Llama~3.1~8B with 32-layer architecture. Unlike the original ICR, which aggregates attention weights across all layers, Selective-ICR aggregates attention weights from a peak-anchored, center-biased interval to form token-level query scores ($s_Q$). The final relevance score ($s$) is the sum of calibrated scores ($s_Q - s_{cal}$) across all document tokens.}
\label{fig:selective-icr}
\end{figure*}

Based on the above observations, we define a unified interval-selection strategy, termed \textit{Center-Biased Interval Selection}, for aggregating layers within the \textit{Interval} ($\mathcal{I}$). This strategy allows the model to preserve peak-centered relevance signals while mitigating dilution from low-informative boundary layers and maintaining cross-domain generalizability, and is governed by the following two constraints:

\begin{itemize}[leftmargin=*, itemsep=2pt]
    \item \updated{\textbf{Peak-Layer Anchoring:} The aggregation interval $\mathcal{I}$} must include the highest-performing layer identified in the pilot study, ensuring that the most informative relevance signal is preserved.
    \item \updated{\textbf{Central Bias Constraint:} Motivated by the observed bell-shaped distribution of relevance signals across transformer layers, the interval is anchored at the peak-performing layer and extended toward the model’s geometric center \updated{to aggregate \textit{neighboring layers}}. While this constraint may result in asymmetric intervals \updated{(i.e., expansion occurs only on one side of peak)}, it prevents the aggregation window from drifting toward boundary layers. In practice, if the peak appears in earlier layers, it defines the lower bound; if it appears in later layers, it defines the upper bound.}
\end{itemize}

\begin{table}[t!]
\setlength{\tabcolsep}{4pt}
\centering
\caption{Pilot study results on TREC-DL datasets. Scores represent the performance of \textbf{all-layer aggregation} (baseline). The \textit{Peak Index} and \textit{Aggregation Interval} ($\mathcal{I}$) illustrate the discovery of the relevance distribution \updated{based on Center-Biased Interval Selection, which will be} used for Selective-ICR.}
\label{tab:pilot_study_results}
\small
\begin{tabular}{lccccccc}
\toprule
\textbf{Model} & \textbf{DL-19} & \textbf{DL-20} & \textbf{Avg.} & \makecell[c]{\textbf{Peak} \\ \textbf{Index}} & \makecell[c]{\textbf{Total} \\ \textbf{Layers}} & \makecell[c]{\textbf{Aggregation} \\ \textbf{Interval }($\mathcal{I}$)} \\
\midrule
BM25 & 0.506 & 0.480 & 0.493 & -- & -- & -- \\
\midrule
Llama 3.1 8B & 0.571 & 0.528 & 0.550 & 18 & 32 & [15, 18] \\
Mistral 7B   & 0.585 & 0.532 & 0.559 & 16 & 32 & [13, 16] \\
Qwen3 0.6B   & 0.491 & 0.439 & 0.465 & 10 & 28 & [10, 13] \\
Qwen3 4B     & 0.545 & 0.490 & 0.517 & 18/21$^*$ & 36 & [18, 21] \\
Qwen3 8B     & 0.508 & 0.470 & 0.489 & 18 & 36 & [18, 21] \\
\bottomrule
\addlinespace[1ex]
\multicolumn{7}{l}{\footnotesize Note: All scores are based on the standard all-layer aggregation protocol \cite{chen:AttentionLargeLanguage:2025}.} \\
\multicolumn{7}{l}{\footnotesize $^*$ Peak index shifted between datasets (18 for DL-19; 21 for DL-20).}
\end{tabular}
\end{table}

Table~\ref{tab:pilot_study_results} summarizes the resulting aggregation intervals ($\mathcal{I}$) for each model. The window size is treated as a configurable parameter~$w$. We instantiate this strategy with a compact window of $w=4$ consecutive layers\textemdash comprising the peak layer and its three neighboring layers\textemdash to enhance cross-domain generalizability (Figure~\ref{fig:selective-icr}). Larger windows may be explored to assess the impact of interval width on robustness and effectiveness. Alternative designs, such as symmetric windows, non-consecutive layer selection based on relevance thresholds, or top-$n$ layer aggregation, are left for future work. All experimental configurations strictly adhere to the two constraints above.

\smallskip
\emph{\textbf{Interval Instantiation.}} Specifically, for the 32-layer architectures (0-indexed, $L \in [0, 31]$), where the peaks occur at indices 18 (Llama~3.1~8B) and 16 (Mistral~7B), we select the 4-layer intervals $[15, 18]$ and $[13, 16]$, respectively.

For the 28-layer Qwen3~0.6B, the signal peaks earlier at index 10. We define its interval as $[10, 13]$ to align with our \textit{Central Bias Constraint}. By shifting the window slightly toward the deeper layers, we position the aggregation closer to the model's geometric center, further isolating the signal from lexical noise in the initial layers.

For the Qwen3~4B and 8B architectures (36 layers), peak performance varies slightly between indices 18 and 21 across different datasets. Accordingly, we select the interval $[18, 21]$ to ensure stable coverage of this high-signal region and also align with the uniform $w=4$ window size. This localized aggregation method, which we term \textit{Selective-ICR}, is employed for all subsequent evaluations on the BEIR benchmark to optimize the trade-off between ranking effectiveness and computational efficiency.

\subsection{Full-Scale Evaluation on BEIR}
We evaluate three ICR-based settings across nine diverse datasets from the BEIR benchmark: the original all-layer ICR (\textit{All}) from Chen et al.~\cite{chen:AttentionLargeLanguage:2025}, a single peak-indexed layer (\textit{Peak}), and the proposed selective interval aggregation using Center-Biased Interval Selection with window size $w=4$ (\textit{Selective}). This evaluation assesses the generalizability of our pilot findings and compares ranking effectiveness and computational efficiency across these settings.

\begin{table*}[t!]
\centering
\caption{Comprehensive comparison of ranking performance (\textit{nDCG@10}) on BEIR datasets across three ICR-based settings: all-layer aggregation (\textit{All}), selective interval aggregation (\textit{Selective}, $[l_1\!-\!l_2]$), and single peak-layer aggregation (\textit{Peak}, $\{l\}$). The best result within each model triad is highlighted in \textbf{bold}. Significance is assessed via per-query paired t-test, comparing \textit{Selective} and \textit{Peak} configurations against \textit{All} baselines ($^{*}$ for $p < 0.05$). The Avg. column reflects macro-average over nine datasets.}
\label{tab:beir_comprehensive}
\resizebox{\textwidth}{!}{
\footnotesize
\begin{NiceTabular}{l|lllllllll|l}
\toprule
\multicolumn{1}{l}{\textbf{Model}} & \textbf{T-Covid} & \textbf{NFCorpus} & \textbf{DBPedia} & \textbf{SciFact} & \textbf{SciDocs} & \textbf{FiQA} & \textbf{FEVER} & \textbf{C-FEVER} & \multicolumn{1}{c}{\textbf{NQ}} & \multicolumn{1}{c}{\textbf{Avg.}} \\
\midrule
BM25 (Baseline) & 0.595 & 0.322 & 0.318 & 0.679 & 0.149 & 0.236 & 0.651 & 0.165 & 0.306 & 0.380 \\
\midrule
\multicolumn{1}{l}{\emph{\textbf{Llama 3.1 8B}}} & \multicolumn{9}{c|}{} & \multicolumn{1}{c}{} \\
\quad All-ICR & 0.728 & 0.347 & 0.353 & \textbf{0.762} & 0.171 & 0.380 & \textbf{0.845} & \textbf{0.232} & 0.512 & 0.481 \\
\quad Selective-ICR, [15--18] & 0.737 & \textbf{0.349} & 0.359 & 0.755 & \textbf{0.172} & \textbf{0.388}~$^{*}$ & 0.841~$^{*}$ & 0.220~$^{*}$ & \textbf{0.522}~$^{*}$ & \textbf{0.483} \\
\quad Peak-ICR, \{18\} & \textbf{0.745} & 0.346 & \textbf{0.366}~$^{*}$ & 0.746~$^{*}$ & 0.169 & 0.366~$^{*}$ & 0.836~$^{*}$ & 0.226~$^{*}$ & 0.520~$^{*}$ & 0.480 \\
\midrule
\multicolumn{1}{l}{\emph{\textbf{Mistral 7B}}} & \multicolumn{9}{c|}{} & \multicolumn{1}{c}{} \\
\quad All-ICR & 0.639 & \textbf{0.331} & 0.314 & 0.724 & \textbf{0.162} & 0.310 & 0.798 & 0.206 & 0.464 & 0.439 \\
\quad Selective-ICR, [13-16] & \textbf{0.650} & \textbf{0.331} & \textbf{0.339}~$^{*}$ & 0.727 & \textbf{0.162} & 0.320~$^{*}$ & \textbf{0.802}~$^{*}$ & \textbf{0.216}~$^{*}$ & 0.484~$^{*}$ & \textbf{0.448} \\
\quad Peak-ICR, \{16\} & 0.636 & 0.328 & 0.337~$^{*}$ & \textbf{0.729} & 0.161 & \textbf{0.326}~$^{*}$ & 0.801 & 0.209 & \textbf{0.487}~$^{*}$ & 0.446 \\
\midrule
\multicolumn{1}{l}{\emph{\textbf{Qwen3}}} & \multicolumn{9}{c|}{} & \multicolumn{1}{c}{} \\
\quad 0.6B All-ICR & 0.614 & 0.325 & \textbf{0.331} & \textbf{0.705} & \textbf{0.155} & \textbf{0.276} & \textbf{0.767} & \textbf{0.212} & \textbf{0.375} & \textbf{0.418} \\
\quad 0.6B Selective-ICR, [10-13] & \textbf{0.627} & \textbf{0.327} & 0.319~$^{*}$ & 0.702 & 0.151~$^{*}$ & 0.272 & 0.759~$^{*}$ & 0.211 & 0.372 & 0.416 \\
\quad 0.6B Peak-ICR, \{10\} & 0.586 & 0.313~$^{*}$ & 0.305~$^{*}$ & 0.665~$^{*}$ & 0.136~$^{*}$ & 0.223~$^{*}$ & 0.661~$^{*}$ & 0.169~$^{*}$ & 0.296~$^{*}$ & 0.373 \\
\midrule
\quad 4B All-ICR  & 0.693 & 0.336 & 0.334 & 0.725 & 0.161 & 0.318 & 0.818 & 0.208 & 0.445 & 0.449 \\
\quad 4B Selective-ICR, [18-21] & \textbf{0.765}~$^{*}$ & \textbf{0.342} & \textbf{0.337} & \textbf{0.741}~$^{*}$ & \textbf{0.166}~$^{*}$ & \textbf{0.355}~$^{*}$ & \textbf{0.837}~$^{*}$ & \textbf{0.222}~$^{*}$ & \textbf{0.496}~$^{*}$ & \textbf{0.473} \\
\quad 4B Peak-ICR, \{18\} & 0.750~$^{*}$ & 0.323~$^{*}$ & 0.327 & 0.726 & 0.162 & 0.326 & 0.824~$^{*}$ & 0.207 & 0.464~$^{*}$ & 0.457 \\
\midrule
\quad 8B All-ICR & 0.693 & 0.337 & 0.326 & 0.724 & 0.159 & 0.321 & 0.820 & 0.197 & 0.448 & 0.447 \\
\quad 8B Selective-ICR, [18-21] & 0.751~$^{*}$ & \textbf{0.342} & \textbf{0.336}~$^{*}$ & \textbf{0.736} & \textbf{0.164}~$^{*}$ & \textbf{0.362}~$^{*}$ & \textbf{0.834}~$^{*}$ & \textbf{0.217}~$^{*}$ & \textbf{0.496}~$^{*}$ & \textbf{0.471} \\
\quad 8B Peak-ICR, \{18\} & \textbf{0.752}~$^{*}$ & 0.327~$^{*}$ & 0.313~$^{*}$ & 0.731 & \textbf{0.164}~$^{*}$ & 0.337~$^{*}$ & 0.821 & 0.202 & 0.469~$^{*}$ & 0.457 \\
\bottomrule
\end{NiceTabular}
}
\vspace{2pt}
\end{table*}

\subsubsection{Ranking Effectiveness}\mbox{} \\
Table~\ref{tab:beir_comprehensive} reports ranking effectiveness for the three ICR-based settings defined above. The results show that isolating intermediate-layer signals preserves or improves ranking precision across diverse domains.

\ready{Overall, four of the five evaluated models achieve their highest average performance under Selective-ICR. Although Peak-ICR performs competitively on specific sub-datasets, it is generally surpassed by selective interval aggregation, indicating that aggregating a small neighborhood of high-signal layers provides greater robustness and generalization than relying on a single isolated layer.}

The most substantial gains are observed in the larger Qwen3~4B and 8B models, where the average $nDCG@10$ improves from approximately $0.449$ to over $0.473$ under Selective-ICR. This significant margin suggests that deeper models are more susceptible to \textit{signal dilution} from boundary layers, which our strategy effectively mitigates. A similar, though more moderate, upward trend is observed in Mistral 7B ($0.439$ to $0.448$) and Llama 3.1 8B ($0.481$ to $0.483$). While the latter shows marginal average gains, Selective-ICR achieves superior results on the majority of tasks, including T-Covid, NFCorpus, SciDocs, FiQA, and NQ, validating the cross-domain consistency of the identified bell-curve distribution. 

In contrast, the smallest model, Qwen3~0.6B, exhibits a slight performance decline under Selective-ICR ($0.418$ to $0.416$), while Peak-ICR performs substantially worse. This result indicates that for ultra-lightweight models, the optimal relevance interval identified on TREC-DL does not generalize well to the diverse domains of BEIR. We speculate that as model depth decreases, the relevance distribution becomes highly task-dependent; consequently, a static interval based on a single benchmark lacks the robustness to capture the shifting signal peaks across different datasets. For such small-scale models, using all layers remains the more reliable strategy.

\subsubsection{Computational Efficiency} \mbox{} \\
Beyond ranking effectiveness, the primary advantage of Selective-ICR lies in its reduction of system latency. In this efficiency analysis, we compare Selective-ICR directly against the original All-ICR baseline~\cite{chen:AttentionLargeLanguage:2025}. Since the computational cost of a transformer forward pass scales linearly with the number of executed layers, terminating inference at the upper bound of the selected interval ($\mathcal{I}_{\max}$) allows the remaining layers to be skipped entirely. We evaluate efficiency along two dimensions: (1) \emph{Forward Pass Latency}, isolating core model inference; and (2) \emph{Total Scoring Latency}, measuring the end-to-end pipeline.

\begin{figure}[t]
    \centering
    \includegraphics[width=\columnwidth]{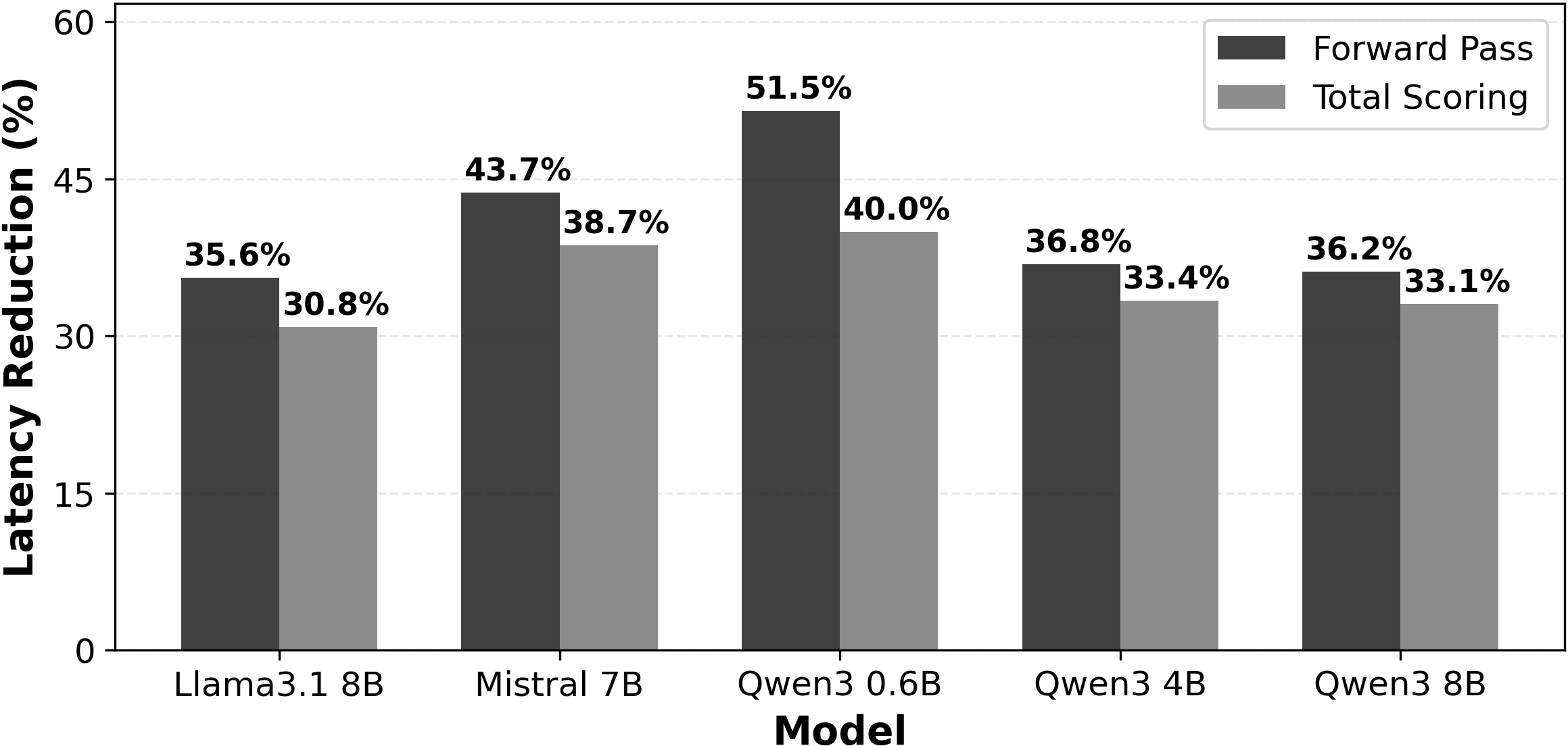}
    \caption{Efficiency gains of the Selective-ICR strategy, measured as the percentage reduction in latency relative to All-ICR for the forward pass and the total scoring stage.}
  \label{fig:timing_comparison}
  \vspace{-0.6cm}
\end{figure}

As illustrated in Figure~\ref{fig:timing_comparison}, the efficiency gains are directly proportional to the reduction in layer depth. For the Forward Pass, restricting the execution to the identified signal-rich regions yields a reduction in inference time, ranging from 35.6\% to 51.5\%. This is most pronounced in the Qwen3 0.6B model; by terminating the forward pass after the first 14 layers (out of 28), the latency is essentially halved. 

Regarding Total Scoring Latency, we observe a consistent decrease of at least 30.8\% across all model families. While this end-to-end gain is slightly lower than the isolated forward pass reduction due to fixed overheads such as tokenization, KV-cache preparation, and attention weight extraction, it nonetheless represents a significant throughput improvement.

\subsubsection{From Empirical Observation to Intrinsic Property} \mbox{} \\
In summary, Selective-ICR shifts the Pareto frontier of zero-shot re-ranking by truncating inference to the model's signal-rich core, achieving both faster execution and improved accuracy over all-layer aggregation.

\begin{figure*}[t]
    \centering
    \includegraphics[width=1.00\textwidth]{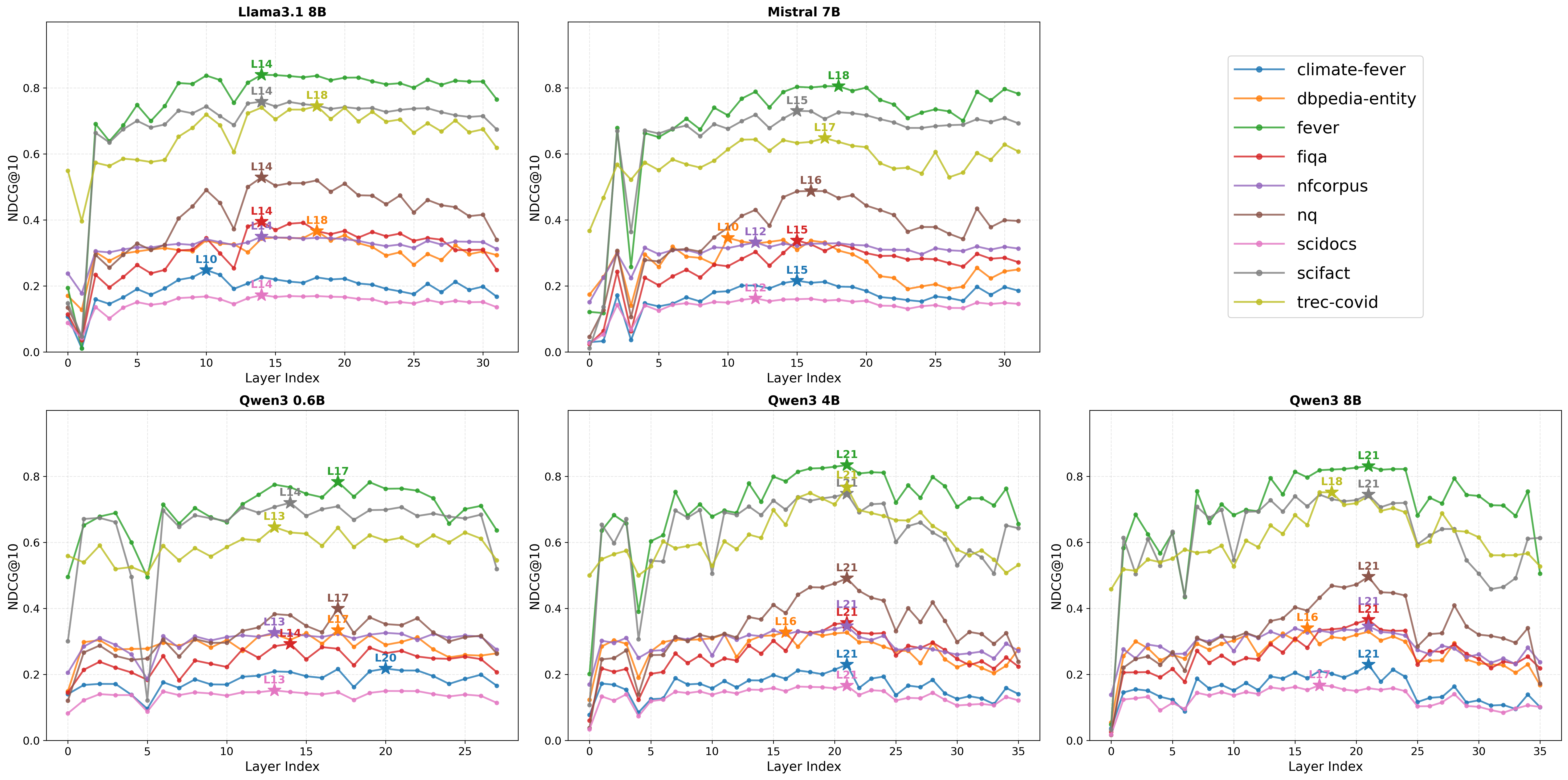}
    \caption{Detailed layer-wise \textit{nDCG@10} trends across diverse BEIR tasks. Each sub-figure represents a specific model's performance trajectory across multiple datasets. Despite the varying semantic domains (e.g., medical, financial, or general knowledge), each model exhibits a highly consistent bell-shaped relevance distribution, peaking within the identified middle-layer intervals.}
    \label{fig:app_full_trends}
\end{figure*}

More importantly, our results indicate that the observed bell-shaped relevance distribution is an intrinsic property of decoder-only architectures rather than a dataset-specific artifact. As shown in Figure~\ref{fig:app_full_trends}, layer-wise performance trends across all BEIR sub-datasets exhibit a highly consistent trajectory: performance rises through early layers, peaks in intermediate layers, and declines toward the final stages. This strong intra-model consistency supports the generality of our findings and validates the design principles underlying Selective-ICR.

%% file: chapter3.tex
\section{Decoupling Ranking and Scoring Strategies}
\label{sec:decoupling}
While our previous analysis established the effectiveness of Selective-ICR as a ranking signal, its relative performance against existing LLM-based scoring approaches within different ranking structures remains unexplored. To address \textbf{RQ2}, this section conducts a controlled decoupling study using Llama~3.1~8B as the backbone model, which achieved the strongest performance under All-ICR and Selective-ICR in our prior experiment. This study also addresses a research direction identified as future work in the original Setwise paper~\cite{zhuang:SetwiseApproachEffective:2024}, which primarily focused on encoder–decoder architectures (T5 series)~\cite{jason:FintunedLanguageModels:2022}.

\begin{table*}[t!]
\centering
\caption{Comprehensive evaluation of Llama 3.1 8B across decoupled ranking frameworks and scoring paradigms. Results are reported for listwise (single-shot and sliding-window) and setwise (heapsort and bubblesort) ranking. Ranks within each subgroup (superscripts) are assigned based on descending \textit{nDCG@10} and ascending latency. IA-based methods are omitted for setwise bubblesort due to a fundamental mismatch in computational requirements for leveraging internal attention.}

\label{tab:rq2_cross_paradigm_results}
\resizebox{\textwidth}{!}{
\begin{NiceTabular}{l|cccccccc|cc}
\toprule
\textbf{Methods} & \textbf{Covid} & \textbf{NFCorpus} & \textbf{Touche} & \textbf{DBPedia} & \textbf{SciFact} & \textbf{Signal} & \textbf{News} & \textbf{Robust04} & \textbf{Avg. \textit{nDCG@10}} & \textbf{Avg. Time (s)} \\
\toprule
BM25 (Baseline) & 0.5950 & 0.3220 & 0.4420 & 0.3180 & 0.6790 & 0.3310 & 0.3950 & 0.4070 & 0.4361 & - \\
\hline
\textbf{\textit{Listwise (window size = 100)}} & & & & & & & & & & \\
\quad \textit{Generation} & 0.6401 & 0.3186 & 0.4094 & 0.3779 & 0.5414 & 0.3090 & 0.3939 & 0.4346 & 0.4281$^{3}$ & 11.6913$^{5}$ \\
\quad \textit{Likelihood} & 0.5548 & 0.3010 & 0.2402 & 0.3130 & 0.6022 & 0.2775 & 0.3078 & 0.3233 & 0.3650$^{5}$ & 1.9586$^{1}$ \\
\quad \textit{Internal Attention} & & & & & & & & & & \\
\qquad All-ICR & 0.6957 & 0.3401 & 0.3567 & 0.3470 & 0.7303 & 0.2840 & 0.4430 & 0.4596 & 0.4571$^{1}$ & 4.3313$^{4}$ \\
\qquad Selective-ICR & 0.7121 & 0.3399 & 0.3217 & 0.3363 & 0.7354 & 0.2707 & 0.4433 & 0.4680 & 0.4534$^{2}$ & 3.5010$^{3}$ \\
\qquad Selective-ICR \& w/o Calibration & 0.6172 & 0.2872 & 0.2853 & 0.2603 & 0.7224 & 0.2205 & 0.3712 & 0.3761 & 0.3925$^{4}$ & 2.3480$^{2}$ \\
\midrule
\textbf{\textit{Listwise (window size = 20)}} & & & & & & & & & & \\
\quad \textit{Generation} & 0.7372 & 0.3413 & 0.3211 & 0.3981 & 0.5624 & 0.3041 & 0.4594 & 0.4969 & 0.4526$^{1}$ & 22.7809$^{5}$ \\
\quad \textit{Likelihood} & 0.7425 & 0.3371 & 0.2707 & 0.3667 & 0.6226 & 0.3084 & 0.4238 & 0.4466 & 0.4398$^{3}$ & 2.9430$^{1}$ \\
\quad \textit{Internal Attention} & & & & & & & & & & \\
\qquad All-ICR & 0.6857 & 0.3344 & 0.2960 & 0.3440 & 0.7135 & 0.2684 & 0.4159 & 0.4252 & 0.4354$^{4}$ & 6.2841$^{4}$ \\
\qquad Selective-ICR & 0.7235 & 0.3397 & 0.2778 & 0.3495 & 0.7291 & 0.2680 & 0.4159 & 0.4486 & 0.4440$^{2}$ & 4.9731$^{3}$ \\
\qquad Selective-ICR \& w/o Calibration & 0.6968 & 0.2906 & 0.2720 & 0.2763 & 0.7085 & 0.2219 & 0.3706 & 0.4077 & 0.4056$^{5}$ & 3.2635$^{2}$ \\
\midrule
\textbf{\textit{Setwise (Heapsort)}} & & & & & & & & & & \\
\quad \textit{Generation} & 0.8140 & 0.3363 & 0.2460 & 0.3770 & 0.7224 & 0.3014 & 0.4662 & 0.5358 & 0.4749$^{1}$ & 14.1067$^{2}$ \\
\quad \textit{Likelihood} & 0.7956 & 0.3342 & 0.2688 & 0.3636 & 0.7201 & 0.2869 & 0.4458 & 0.5170 & 0.4665$^{2}$ & 8.8550$^{1}$ \\
\quad \textit{Internal Attention} & & & & & & & & & & \\
\qquad All-ICR & 0.7209 & 0.3510 & 0.2946 & 0.3632 & 0.7358 & 0.3047 & 0.4635 & 0.4570 & 0.4613$^{4}$ & 42.1332$^{5}$ \\
\qquad Selective-ICR & 0.7277 & 0.3549 & 0.2622 & 0.3752 & 0.7448 & 0.2958 & 0.4737 & 0.4884 & 0.4653$^{3}$ & 36.0134$^{4}$ \\
\qquad Selective-ICR \& w/o Calibration & 0.7208 & 0.3290 & 0.2675 & 0.2622 & 0.7405 & 0.2429 & 0.4790 & 0.4752 & 0.4396$^{5}$ & 17.7519$^{3}$ \\
\midrule
\textbf{\textit{Setwise (Bubblesort)}} & & & & & & & & & & \\
\quad \textit{Generation} & 0.8023 & 0.3547 & 0.3222 & 0.3910 & 0.7394 & 0.3411 & 0.4762 & 0.5466 & 0.4967$^{2}$ & 50.9419$^{2}$ \\
\quad \textit{Likelihood} & 0.7972 & 0.3500 & 0.3451 & 0.3931 & 0.7340 & 0.3491 & 0.4905 & 0.5404 & 0.4999$^{1}$ & 31.5448$^{1}$ \\
\bottomrule
\end{NiceTabular}
}
\end{table*}

\subsection{Experimentation}
\subsubsection{Ranking Frameworks}\mbox{} \\
\vspace{-0.2cm}
\begin{itemize}[leftmargin=*, itemsep=2pt, topsep=0pt]
    \item \textbf{Listwise (Sliding Window):} This framework processes a list of candidate documents within a single prompt, typically relying on the model's ability to generate a full ranked list of identifiers~\cite{zhuang:SetwiseApproachEffective:2024}. To accommodate model context limits, a sliding window mechanism is employed to iteratively re-rank sub-sequences, progressing from the bottom of the initial ranking upwards.
    \item \textbf{Setwise (Heapsort / Bubblesort):} This approach decomposes the ranking task into a series of set-based comparisons, where the LLM selects the most relevant document from a candidate set of size $c$. By increasing the set size ($c > 2$), the method significantly reduces the total number of LLM inferences required for re-ranking. We implement setwise variants of Heapsort ($O(k \cdot \log_c N)$) and Bubblesort ($O(k \cdot N/(c-1))$) to efficiently identify the top-$k$ documents from a pool of $N$ candidates.
\end{itemize}

\subsubsection{Scoring Mechanisms}\mbox{} \\
\begin{itemize}[leftmargin=*, itemsep=2pt, topsep=0pt]
    \item \textbf{Generation (Gen):} This paradigm relies on the model's ability to predict the next token to produce document identifiers. In listwise ranking, the LLM is prompted to generate a complete ordered list of document labels (e.g., $[2] > [3] > [5] \dots$), while in setwise ranking, it generates the identifier of the most relevant document from a given candidate set.

    \item \textbf{Likelihood (Lik):} This method leverages the model's output logits to simulate the probability of a specific document label being the ``most relevant'' among a set. Regardless of the ranking structure, the LLM is always prompted to identify only the single most relevant candidate. In listwise ranking, candidates are sorted globally by these label probabilities; in setwise ranking, the highest probability (argmax) is used to select a single winner within each local comparison.

    \item \textbf{Internal Attention (IA):} Following the ICR framework~\cite{chen:AttentionLargeLanguage:2025}, this mechanism derives relevance scores directly from the attention weights between query and document tokens across the model's hidden layers. While a compact aggregation window ($w=4$) is sufficient to capture the dominant relevance signal around the peak layer, we observe that the peak position can vary mildly across datasets for Llama~3.1~8B. To improve robustness under such variations, we apply the Selective-ICR strategy with a wider, symmetric window of $w=6$ layers ($L \in [13,18]$) in the decoupling study. This design preserves coverage of the signal-rich core while mitigating sensitivity to small peak shifts, enabling more stable performance across iterative ranking settings.
\end{itemize}

\subsubsection{Experimental Setting}\mbox{} \\
To ensure a rigorous and comparable evaluation, we follow the experimental configurations established in the Setwise paper~\cite{zhuang:SetwiseApproachEffective:2024}. Specifically, despite the much larger context window offered by Llama~3.1, we maintain the maximum document length at 128 tokens and set the number of compared documents per step to $c=3$ (comprising one parent and two child nodes) for the Setwise approach. We conducted all experiments on an NVIDIA L40 (48GB) GPU.

Furthermore, we introduce an ablation variant for the attention scorer: \textit{w/o Calibration}. While the standard ICR requires a dual-pass approach (one for the query and one for a ``N/A'' content-free prompt) to mitigate intrinsic model bias, the w/o Calibration version utilizes only a single forward pass. This design allows us to explicitly measure the trade-off between the precision gains afforded by the ``N/A'' prompt calibration and the increased throughput of a single-inference setting.

\subsection{Listwise Comparative Analysis}
The evaluation of listwise ranking, as summarized in Table~\ref{tab:rq2_cross_paradigm_results}, reveals clear performance differences among scoring paradigms under different context windowing strategies.
Under single-shot listwise inference (window size = 100), \textit{internal attention} (IA) exhibits clear superiority, achieving an $nDCG@10$ of 0.4571 for All-ICR, followed closely by Selective-ICR at 0.4534 with only a marginal reduction in latency. Both substantially exceed the results of \textit{generation}-based scoring (Gen, 0.4281) and \textit{likelihood}-based scoring (Lik, 0.3650).

In contrast, the performance rankings shift markedly when adopting a sliding-window strategy (window size = 20, step = 10). While this iterative approach allows Gen and Lik scoring to improve over their single-shot performance from 0.4281 and 0.3650 to 0.4526 and 0.4398 respectively, IA generally struggles with the loss of global context. Specifically, the baseline All-ICR experiences a performance drop from 0.4571 to 0.4354, while its latency climbs to 6.28s due to the overhead of multiple forward passes. Importantly, Selective-ICR remains robust under this constrained setting, achieving an $nDCG@10$ of 0.4440\textemdash ranking second overall and slightly below Gen (0.4526)\textemdash while maintaining a lower inference time (4.97s vs.\ 22.78s). This robustness is particularly relevant in practice, as the effective input context of an LLM is bounded by its architectural context limit, constrained by GPU memory and inference configuration, and further shaped by task-specific prompt and ranking design. As a result, ranking frameworks frequently operate under restricted context windows, even for long-context models.

While Selective-ICR remains competitive, the overall downward trend suggests that IA-based signals fundamentally rely on all-to-all attention over the full candidate set; when restricted to local windows, their global discriminative power diminishes, making them less compatible with iterative refinement strategies that typically favor generative and likelihood-based rankers.

\subsection{Setwise Comparative Analysis}
Moving to Setwise frameworks exposes the inherent limitations of internal attention signals when restricted to local contexts. Overall, Setwise frameworks (Heapsort/Bubblesort) exhibit generally higher effectiveness compared to their corresponding listwise approaches. This gain is attributed to the iterative nature of setwise logic: each comparison operates on a small window ($c=3$), and the $O(N \log N)$ or $O(N^2)$ sequence of fine-grained updates enables progressively refined ranking decisions. Furthermore, the relative performance ordering of scoring paradigms remains consistent with the trend observed when reducing the listwise window size (from $w=100$ to $w=20$), confirming that IA-based signals are less effective when restricted to a limited candidate set.

However, this gain in effectiveness under Setwise frameworks comes at a high computational cost, particularly for internal attention–based methods. Unlike listwise ranking, which performs a single forward pass comparison over all $N$ documents, setwise frameworks require at least $O(N \log N)$ forward passes over small candidate subsets. Although increasing the setwise capacity $c$ reduces the number of forward passes, our trials show no net efficiency gain; the increased latency per comparison (due to richer attention extraction) offsets the fewer calls.

Theoretically, removing calibration should allow IA to outperform generative methods, as it bypasses the decoding stage. However, our ablation study reveals that even without calibration, Selective-ICR ($17.75s$) remains slower than the generative baseline ($14.11s$). We attribute this gap to the heavy overhead of processing internal signals. While Gen/Lik scoring only requires extracting the final-layer output or logits, IA-based methods must capture and store large attention matrices in GPU memory from multiple layers. The extra work of managing these massive tensors and calculating the attention scores further outweighs the benefit of skipping token decoding.

Bubblesort represents the extreme of these trends. While Lik-scoring achieves peak effectiveness ($0.4999$ $nDCG@10$), slightly outperforming Gen ($0.4967$), IA variants are omitted. Their per-comparison overhead becomes prohibitive under the large number of fine-grained swaps required by bubblesort, leading to impractical runtimes without corresponding gains. 

Overall, a clear specialization emerges from the results in Table~\ref{tab:rq2_cross_paradigm_results}: IA-based scoring is best suited for high-throughput, global ranking (e.g., $w=100$), whereas explicit output-based methods (Gen/Lik) are the superior choice for localized, fine-grained refinement where iterative updates are required.

%% file: chapter4.tex
\section{Robustness across Reasoning Domains}
\label{sec:rq3}
We address \textbf{RQ3} by evaluating whether \textit{internal attention} (IA) signals effectively capture relevance in tasks requiring complex query reasoning and deep understanding. This section examines if these internal signals remain reliable when retrieval depends on logical deduction rather than simple semantic or keyword matching.

\subsection{Experimental Setup on BRIGHT}
To evaluate whether internal attention (IA) signals remain effective in reasoning-intensive retrieval, we use the BRIGHT benchmark~\cite{su:BRIGHTRealisticChallenging:2025}. BRIGHT contains 1,385 real-world queries spanning diverse domains such as StackExchange, LeetCode, and mathematics competitions, where retrieval depends on deliberate logical reasoning rather than surface-level semantic matching.

We re-rank the top-100 BM25 candidates using five decoder-only models: Llama~3.1~8B, Mistral~7B, and Qwen3 (0.6B, 4B, and 8B). We evaluate two IA-based configurations:
\begin{enumerate}[font=\bfseries,leftmargin=0.2cm, itemindent=0.8cm]
    \item \textbf{All-ICR:} identical to the all-layer setting in Section~\ref{sec:rq1}; and 
    \item \textbf{Dataset-Specific Peak-ICR (Oracle):} relevance derived from the single best-performing layer identified \emph{per dataset} via layer-wise evaluation on BRIGHT itself. Unlike the cross-dataset interval selection used in earlier sections, this configuration selects the optimal layer separately for each BRIGHT sub-task, providing an upper bound on single-layer IA performance.
\end{enumerate}

We report $nDCG@10$ as the primary metric and compare against competitive reasoning-focused baselines, including RankZephyr-7B, a state-of-the-art listwise reranker~\cite{pradeep:RankZephyrEffectiveRobust:2023}; RankGPT-4, a zero-shot listwise reranker where GPT-4~\cite{openai:GPT4TechnicalReport:2024} is used as backbone; and Rank-R1 variants~\cite{zhuang:RankR1EnhancingReasoning:2025}, which utilize explicit reinforcement learning (GRPO: Group Relative Policy Optimization) to enhance reasoning logic.

\begin{table*}[t!]
\centering
\caption{Re-ranking effectiveness (\textit{nDCG@10}) on the BRIGHT benchmark. We compare All-ICR and Oracle Peak-ICR against competitive reasoning-focused baselines. $^{*}$ Results are reported from~\cite{zhuang:RankR1EnhancingReasoning:2025,su:BRIGHTRealisticChallenging:2025}.}
\label{tab:rq3_bright_results}
\resizebox{\textwidth}{!}{
\begin{NiceTabular}{l|cccccccccccc|c}
\toprule
\textbf{Model} & \textbf{AoPS} & \textbf{Bio.} & \textbf{Earth.} & \textbf{Econ.} & \textbf{Leet.} & \textbf{Pony} & \textbf{Psy.} & \textbf{Rob.} & \textbf{Stack.} & \textbf{Sus.} & \textbf{TheoQ.} & \textbf{TheoT.} & \textbf{Avg.} \\
\midrule
BM25 & 0.0645 & 0.1810 & 0.2808 & 0.1646 & 0.2473 & 0.0434 & 0.1341 & 0.1092 & 0.1619 & 0.1613 & 0.0734 & 0.0213 & 0.1369 \\
\midrule
\textbf{\textit{Competitive Baselines}$^{*}$} & & & & & & & & & & & & & \\
\quad RankZephyr-7B & 0.0680 & 0.2190 & 0.2370 & 0.1440 & 0.2470 & 0.0650 & 0.1030 & 0.0760 & 0.1370 & 0.1660 & 0.0730 & 0.0200 & 0.1300 \\
\quad RankGPT-4 (Zeroshot) & 0.0120 & \textbf{0.3380} & 0.3420 & 0.1670 & 0.0340 & \textbf{0.1560} & \textbf{0.2700} & 0.2230 & \textbf{0.2770} & 0.1110 & 0.0020 & 0.0860 & 0.1700 \\
\quad Rank-R1-7B (GRPO) & 0.0430 & 0.2600 & 0.2850 & 0.1720 & 0.1980 & 0.0430 & 0.2420 & 0.1910 & 0.1040 & 0.2420 & 0.0830 & 0.1090 & 0.1640 \\
\quad Rank-R1-14B (GRPO) & 0.0970 & 0.3120 & 0.3850 & 0.2120 & 0.2020 & 0.0920 & 0.2640 & \textbf{0.2260} & 0.1890 & \textbf{0.2750} & 0.0920 & \textbf{0.1190} & 0.2050 \\
\midrule
\textbf{\textit{All-ICR}} & & & & & & & & & & & & & \\
\quad Llama 3.1 8B & 0.0993 & 0.3011 & 0.3759 & 0.2135 & 0.2183 & 0.0156 & 0.2263 & 0.1756 & 0.1292 & 0.1638 & 0.1189 & 0.0621 & 0.1750 \\
\quad Mistral 7B & 0.0945 & 0.2764 & 0.2881 & 0.1523 & 0.1918 & 0.0063 & 0.1572 & 0.1075 & 0.0467 & 0.1285 & 0.1026 & 0.0575 & 0.1341 \\
\quad Qwen3 0.6B & 0.0921 & 0.2994 & 0.3335 & 0.1510 & 0.2289 & 0.0258 & 0.1710 & 0.1322 & 0.0852 & 0.1615 & 0.0923 & 0.0185 & 0.1493 \\
\quad Qwen3 4B & 0.0747 & 0.2286 & 0.2955 & 0.1145 & 0.1690 & 0.0322 & 0.1756 & 0.1332 & 0.0926 & 0.1406 & 0.0954 & 0.0242 & 0.1313 \\
\quad Qwen3 8B & 0.0718 & 0.2242 & 0.2768 & 0.1286 & 0.1575 & 0.0351 & 0.1770 & 0.1381 & 0.0679 & 0.1289 & 0.0857 & 0.0173 & 0.1258 \\
\midrule
\textbf{\textit{Oracle Peak-ICR}} & & & & & & & & & & & & & \\
\quad Llama 3.1 8B & 0.1080 & 0.3295 & \textbf{0.4150} & \textbf{0.2570} & 0.2887 & 0.0392 & 0.2526 & 0.2214 & 0.1862 & 0.2114 & \textbf{0.1296} & 0.0716 & \textbf{0.2092} \\
\quad Mistral 7B & 0.1130 & 0.3055 & 0.3431 & 0.2064 & 0.2676 & 0.0403 & 0.2143 & 0.1650 & 0.1096 & 0.1712 & 0.1163 & 0.0597 & 0.1760 \\
\quad Qwen3 0.6B & 0.1105 & 0.3093 & 0.3746 & 0.1936 & 0.2785 & 0.0821 & 0.2004 & 0.1846 & 0.1487 & 0.1773 & 0.1148 & 0.0552 & 0.1858 \\
\quad Qwen3 4B & 0.1041 & 0.3268 & 0.4037 & 0.2159 & 0.2883 & 0.0784 & 0.2306 & 0.1993 & 0.1644 & 0.1910 & 0.1180 & 0.0627 & 0.1986 \\
\quad Qwen3 8B & \textbf{0.1167} & 0.3032 & 0.4020 & 0.2101 & \textbf{0.2902} & 0.0807 & 0.2422 & 0.2029 & 0.1670 & 0.2158 & 0.1126 & 0.0560 & 0.1999 \\
\bottomrule
\end{NiceTabular}
}
\end{table*}

\subsection{All-Layer Aggregation Analysis}
The all-layer configuration (All-ICR) serves as the baseline for internal attention on BRIGHT. As shown in Table~\ref{tab:rq3_bright_results}, BM25 achieves an average $nDCG@10$ of 0.1369. Under All-ICR, Llama~3.1~8B reaches 0.1750 on average across BRIGHT subtasks and Qwen3~0.6B attains 0.1493, both outperforming the lexical baseline. In contrast, Mistral~7B (0.1341) and the larger Qwen3 models\textemdash 4B (0.1313) and 8B (0.1258)\textemdash underperform BM25.

Interestingly, the smallest model (0.6B) outperforms its larger Qwen3 variants under All-ICR. We attribute this to a \textit{signal dilution effect}: as model depth increases, aggregating attention across all layers introduces additional noise that can obscure relevance signals. For deeper models, indiscriminate layer aggregation becomes counterproductive, a trend further clarified in the subsequent Oracle Peak-ICR analysis.

\subsection{Theoretical Ceiling: Oracle Peak-ICR}
\label{subsec:best}
Selecting the dataset-specific peak layer (Oracle Peak-ICR) substantially improves performance across models, with gains that scale with model size. As shown in Table~\ref{tab:rq3_bright_results}, Qwen3~0.6B increases from 0.1493 to 0.1858 (+24.45\%), while Qwen3~8B rises from 0.1258 to 0.1999 (+58.90\%). This widening improvement strongly supports the signal dilution hypothesis: larger models contain stronger reasoning signals, but these signals are obscured when all layers are aggregated indiscriminately.

Llama~3.1~8B improves from 0.1750 to 0.2092 (+19.54\%), achieving the highest overall average and surpassing competitive reasoning-focused baselines, including the largest 14B Rank-R1 model. It markedly outperforms BM25 in reasoning-heavy domains such as \textit{Earth Science} (0.4150) and \textit{Economics} (0.2570). Notably, even the Qwen3 0.6B (Oracle Peak-ICR) model reaches 0.1858, surpassing RankZephyr-7B (0.130), RankGPT-4 (0.170), and the RL-trained Rank-R1-7B (0.164) without reasoning-specific fine-tuning. This demonstrates that zero-shot internal attention signals can compete with specialized reasoning models, highlighting their substantial latent potential. Fully exploiting this representational capacity toward the Oracle Peak theoretical ceiling remains an important direction for future work (Section~\ref{sec:conclusion}).

\subsection{Attention Distribution across Reasoning Domains}
\begin{figure*}[t!]
    \centering
    \includegraphics[width=1.00\textwidth]{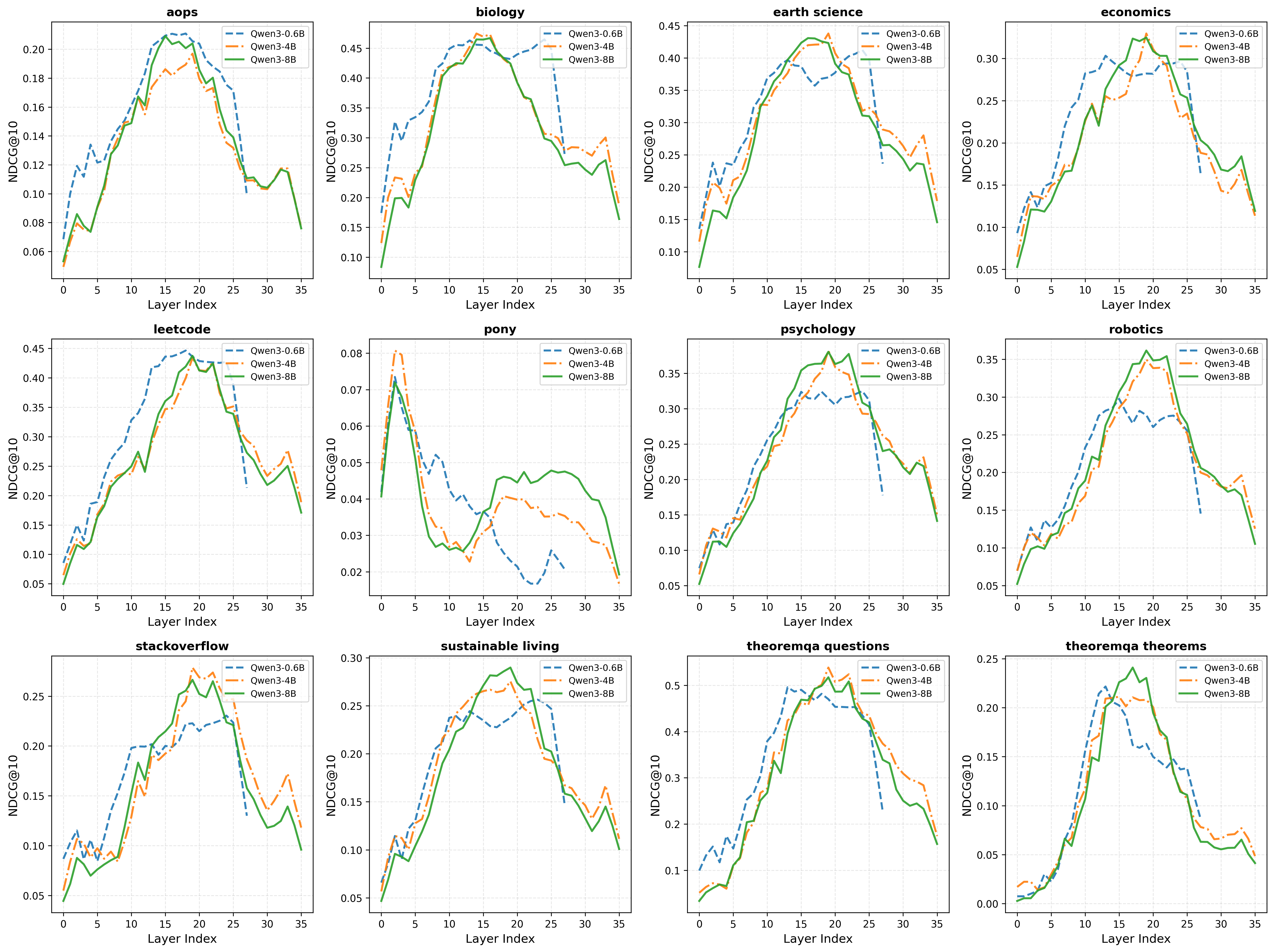}
    \vspace{-0.2cm}
    \caption{Overall distribution of \textit{nDCG@10} scores across the 12 sub-domains of the reasoning-intensive BRIGHT benchmark for Qwen3 variants. The plots reveal a consistent ``bell curve'' trend across scientific/mathematical domains, with peak performance concentrated in intermediate layers, while the programming syntactic primitive retrieval task (\textit{Pony}) exhibits a distinct early-layer signal concentration.}
    \label{fig:bright_layer_distribution}
\end{figure*}

A key finding is that the layer-wise ``bell curve'' distribution appears to be a consistent structural characteristic of decoder-only transformers. This pattern remains remarkably stable even when transitioning from surface-level semantic matching to complex logical reasoning in the BRIGHT benchmark. As illustrated in Figure~\ref{fig:bright_layer_distribution}, this trend holds across the vast majority of domains. To better identify the overall directional trend of the relevance signal, we apply a simple moving average to the visualizations. A notable exception is observed in \textit{Pony}, a rare programming language where target documents consist of formal manuals with dense, rule-based syntax~\cite{su:ARKSActiveRetrieval:2024}. In this specific case, the discriminative signals shift toward earlier layers, likely driven by the increased demand for structural and syntactic parsing inherent in formal programming languages~\cite{yin2025neuron}. The persistence of the peak-layer phenomenon\textemdash even under high reasoning pressure\textemdash supports the viability of peak-oriented layer-selection strategies: Peak-ICR as an oracle when dataset-specific peak layers can be identified, and Selective-ICR as a practical approximation that enables cross-domain generalizability across diverse retrieval tasks.

%% file: chapter5.tex
\section{Conclusion and Future Work}
\label{sec:conclusion}
This paper presents a comprehensive investigation of internal attention mechanisms in Large Language Models (LLMs) for zero-shot document re-ranking. Through extensive evaluation across multiple decoder-only families (Llama~3.1, Mistral, and Qwen3), we establish several key findings.

First, we identify a consistent ``bell-curve'' distribution of relevance signals across transformer layers, with peak effectiveness concentrated in middle-layer intervals and diminished utility in early and late layers. Based on this observation, we propose \textit{Selective-ICR} with \textit{Center-Biased Interval Selection}, which isolates signal-rich layers while truncating unnecessary computation. This strategy preserves or improves ranking effectiveness and reduces inference latency by 30\%--50\% by avoiding non-informative layers.

Second, we conduct the first orthogonal evaluation of scoring mechanisms (Generation, Likelihood, and Internal Attention) across both Listwise and Setwise ranking frameworks. Although ICR provides $O(1)$ complexity in theory, empirical results on Llama~3.1~8B show that dual-pass calibration introduces substantial constant overhead. ICR performs strongly in single-pass, large-context listwise settings but becomes less competitive under iterative sliding-window and setwise configurations, where Likelihood-based scoring often offers superior efficiency-effectiveness trade-offs. These findings position ICR as particularly suitable for high-throughput, large-context ranking rather than fine-grained local comparisons.

Finally, our evaluation on the BRIGHT benchmark shows that although standard attention-based aggregation faces limitations in reasoning-intensive tasks, isolating the optimal layer reveals substantial latent effectiveness. A zero-shot Llama~3.1~8B model can achieve performance parity with much larger RL-trained models, while even the small 0.6B variant surpasses state-of-the-art generative rankers such as RankGPT-4. These results highlight the potential of internal attention signals for reasoning-heavy retrieval\textemdash provided that the most informative internal representations are precisely isolated.

Therefore, reaching this theoretical ceiling remains an important direction for future work. Beyond the Center-Biased Interval Selection strategy explored here, future research may investigate alternative layer-selection designs, such as symmetric windows, non-consecutive layer selection based on relevance thresholds, or adaptive top-$n$ aggregation. More fine-grained optimization may further benefit from head-level analysis: recent work on \textit{Query-Focused Retrieval Heads} (QRHeads)~\cite{zhang:QueryFocusedRetrievalHeads:2025} shows that isolating query-responsive heads improves retrieval effectiveness. A unified framework that jointly selects informative layers and attention heads may therefore provide a principled approach to mitigating architectural noise and fully exploiting internal attention signals for reasoning-intensive retrieval. Furthermore, exploring the effectiveness of Selective-ICR within specialized domain agents~\cite{chen2025beyond} remains a promising direction for future research.

%% file: references.tex
\bibliographystyle{ACM-Reference-Format}
\bibliography{ICR.bib}